\documentclass[sigconf]{acmart}

\usepackage{booktabs} 

\usepackage{multicol}
\usepackage{subcaption}
\usepackage{algorithmic}
\usepackage{graphicx}
\usepackage{textcomp}
\usepackage{xcolor}
\usepackage{caption}
\usepackage{float}

\copyrightyear{2023}
\acmYear{2023}
\setcopyright{rightsretained}
\acmConference[SAC '23]{The 38th ACM/SIGAPP Symposium on Applied Computing}{March 27 - March 31, 2023}{Tallinn, Estonia}
\acmBooktitle{The 38th ACM/SIGAPP Symposium on Applied Computing (SAC '23), March 27 - March 31, 2023, Tallinn, Estonia}\acmDOI{10.1145/3555776.3577611}
\acmISBN{978-1-4503-9517-5/23/03}
\begin{document}

\title{Topology Analysis of the XRP Ledger}
  
\renewcommand{\shorttitle}{ Topology Analysis of the XRP Ledger}

\author{Vytautas Tumas}
\authornote{Corresponding author.}
\affiliation{%
  \institution{University of Luxembourg}
  \streetaddress{SEDAN - SnT}
  \city{Luxembourg} 
  \postcode{}
  \state{} 
  \country{Luxembourg}
}
\email{vytautas.tumas@uni.lu}

\author{Sean Rivera}
\affiliation{%
  \institution{University of Luxembourg}
  \streetaddress{}
  \city{Luxembourg} 
  \postcode{}
  \state{} 
  \country{Luxembourg}
}
\email{sean.rivera@uni.lu}
\author{Damien Magoni}
\affiliation{%
  \institution{University of Bordeaux}
  \streetaddress{}
  \city{Talence} 
  \state{} 
  \country{France}
  \postcode{}
}
\email{magoni@labri.fr}
\author{Radu State}
\affiliation{%
  \institution{University of Luxembourg}
  \streetaddress{}
  \city{Luxembourg} 
  \postcode{}
  \state{} 
  \country{Luxembourg}
}
\email{radu.state@uni.lu}


\begin{abstract}
XRP Ledger is one of the oldest, well-established blockchains. Despite the popularity of the XRP Ledger, little is known about its underlying peer-to-peer network. The structural properties of a network impact its efficiency, security and robustness. We aim to close the knowledge gap by providing a detailed analysis of the XRP overlay network.

In this paper we examine the graph-theoretic properties of the XRP Ledger peer-to-peer network and its temporal characteristics. We crawl the XRP Ledger over two months and collect 1,290 unique network snapshots. We uncover a small group of nodes that act as a networking backbone. In addition, we observe a high network churn, with a third of the nodes changing every five days. Our findings have strong implications for the resilience and safety of the XRP Ledger.
\end{abstract}

%
%



\maketitle

\section{Introduction}

XRP Ledger is one of the oldest, well-established cryptocurrencies. In 2022 it ranked seventh by market capitalization. The XRP Ledger aims to provide high transaction throughput whilst maintaining security against Byzantine failures. The XRP Ledger Consensus Protocol is a Federated Byzantine Agreement protocol~\cite{Rebello}, in which each participant selects a Unique Node List (UNL) of validators. These validators are not necessarily trusted individually but are believed not to collude as a collective. 

Servers running the \textit{rippled} software join into a single peer-to-peer network. The peer-to-peer network's topological structure affects the blockchain's security, resilience, and efficiency. By design, there are no direct incentives to run \textit{rippled} software~\cite{xrpIncentive}. Those who participate do so because they are interested in the long-term health of or are participants on the XRP Ledger.  A corpus of research focuses on the study of structural properties of Bitcoin~\cite{miller2015discovering,delgado2018} and Ethereum~\cite{Gao2019,zhao2021,paphitis2021}. To the best of our knowledge, there are no works examining the overlay network of the XRP Ledger.

The overlay network is uniquely suited for study. Unlike other blockchains that focus on hiding their topology, XRP Ledger natively supports network crawling~\cite{xrp:crawl}. The public availability of data enables researchers to determine the accurate topology of the network. 

In this paper, we provide an in-depth analysis of the graph-theoretic properties of the XRP Ledger overlay network. Our main contributions are as follows:
\begin{itemize}
	\item We measure the structural properties of the network, as well as their evolution over time. We discover a central component of the network.
\item We examine the stability of the nodes and their uptime. We show that less than 50\% of nodes maintained their presence during the measurement period.
	\item Finally, we show that the network may be vulnerable to Autonomous System failures.
\end{itemize}

The remainder of this paper is structured as follows. We discuss related work in Section~\ref{sec:related}. In Section~\ref{sec:background}, we introduce the relevant aspects of the XRP network. We describe our findings in Section~\ref{sec:results} and in Section~\ref{sec:temporal-analysis} we discuss network changes over time. Finally, we conclude our work in Section~\ref{sec:discuss}.

\section{Related Work}
\label{sec:related}

We discovered a significant corpus studying cryptocurrency networks, predominantly Bitcoin and Ethereum. We provide a summary of these works in this Section. 

Miller \emph{et al.}~\cite{miller2015discovering} were the first to determine the topology of the Bitcoin network. The authors discovered \textit{"extremely high-degree nodes"}, which persist in the network over time. Furthermore, the Bitcoin network is not purely random. 
Delgado-Segura \emph{et al.}~\cite{delgado2018} inferred the topology of Bitcoin using orphaned transactions. Due to the limitations of their method, they performed measurements only in the Bitcoin \textit{testnet}. Their results indicate that the testnet is not a random graph.

Paphitis et.al.~\cite{paphitis2021} conducted a graph-theoretic analysis of several different blockchain overlay networks. The results indicate that blockchain overlays have varying network properties and degree distributions. Despite the significant variance, there is a strong correlation between the node's session length and the degree. In addition, the networks have small average shortest paths, but they are not small-world. Finally, the overlay networks are resilient to random node failures, but targeted attacks can considerably affect their connectivity.

Similar studies focus on the Ethereum block-chain. Zhao \emph{et al.}~\cite{zhao2021} performed a temporal, evolutionary analysis of the Ethereum blockchain interaction networks. The authors found a link between anomalies in structural properties and real-life events. Furthermore, they discovered that the network expansion follows a preferential attachment model.

In a later study, Gao \emph{et al.}~\cite{Gao2019} conducted a graph-theoretic analysis of the peer-to-peer layer of the Ethereum network. They discovered an abundance of nodes that do not contribute to the Ethereum network. Furthermore, they showed that the degree distribution does not follow a power-law. In contradiction to the work of Paphitis \emph{et al.}, the authors found evidence of small-world property.

The research conducted in the XRP Ledger context is predominantly on the Consensus Protocol. Chase \emph{et al.}~\cite{chase2018analysis} provide a detailed description and analysis of the Consensus Protocol. They demonstrate that at least a 90\% overlap of the UNLs is required to ensure network safety. In a later study, Christodoulou \emph{et al.}~\cite{christodoulou2020} show when fewer than 20\% of nodes are malicious, the overlap of UNLs can be relaxed. Otherwise, an overlap of 90-99\%  is required. In a similar study, Amores-Sesar \emph{et al.}~\cite{amoressesar2020security} demonstrate that, in the presence of Byzantine nodes, the ledger may fork under standard UNL overlap requirements. Furthermore, the authors show that a single Byzantine node may cause consensus protocol to lose liveliness. 

In a different line of research, Roma \emph{et al.}~\cite{roma2020energy} studied the energy efficiency of an XRP validator. They found that the annual validator running cost is significantly lower than that of a miner.

Aoyama\cite{aoyama2021xrp} provides a unique view of the XRP network from the perspective of its transactions. They found a clear divide between groups accepting transactions and groups receiving transactions.

To the best of our knowledge, there are no previous studies on the topological properties of the XRP network. With this work, we aim to close this gap.

\section{Background}
\label{sec:background}

The XRP Ledger consists of nodes running the \textit{rippled}~\cite{xrp:source-code} software. The interconnected \textit{rippled} servers form the decentralized peer-to-peer overlay network. 

The node owners configure it to accept some number of inbound and outbound connections. Each outgoing connection corresponds to an incoming connection at another node. When nodes connect, the communication over the link is bidirectional. We, therefore, represent the overlay network as a directed graph. The direction of an edge identifies the node that initiated the connection.

A node gets initial entry into the overlay network by connecting to several hardcoded bootstrapping hubs. These hubs share the addresses of other nodes with available inbound connections. The node continues to establish links to others until it reaches the desired limit of outgoing connections. When a node has reached its maximum number of inbound links, it rejects further connection attempts. At the time of data collection, the default number of outgoing connections was 10.

A node periodically advertises to its peers when it has open inbound connection slots. Nodes store this information and communicate it with their peers. As a result, available incoming connections propagate throughout the network.

\section{Network Analysis}
\label{sec:results}

\paragraph{Network Topology} We used the XRP Ledger Crawler~\cite{xrp:crawler-code} to discover the nodes in the overlay network. The crawler starts by querying the peers of a single \textit{rippled} server\footnote{We used \textit{r.ripple.com} as the starting node}. It adds the new nodes to a list and calls every node with a known IP address. The crawler repeats this process until it no longer discovers new nodes.  We enriched the snapshot data with Autonomous System information. We collected the XRP Ledger network snapshot for two months, between 05/01/2022 and 01/03/2022. We crawled the network topology at one-hour intervals. In two months, we collected 1,290 snapshots. We made the datasets available online for further research~\cite{dataUrl}.

\begin{figure*}[htbp]   
    \begin{subfigure}[t]{0.95\columnwidth}
        \centering 
        \includegraphics[width=\textwidth]{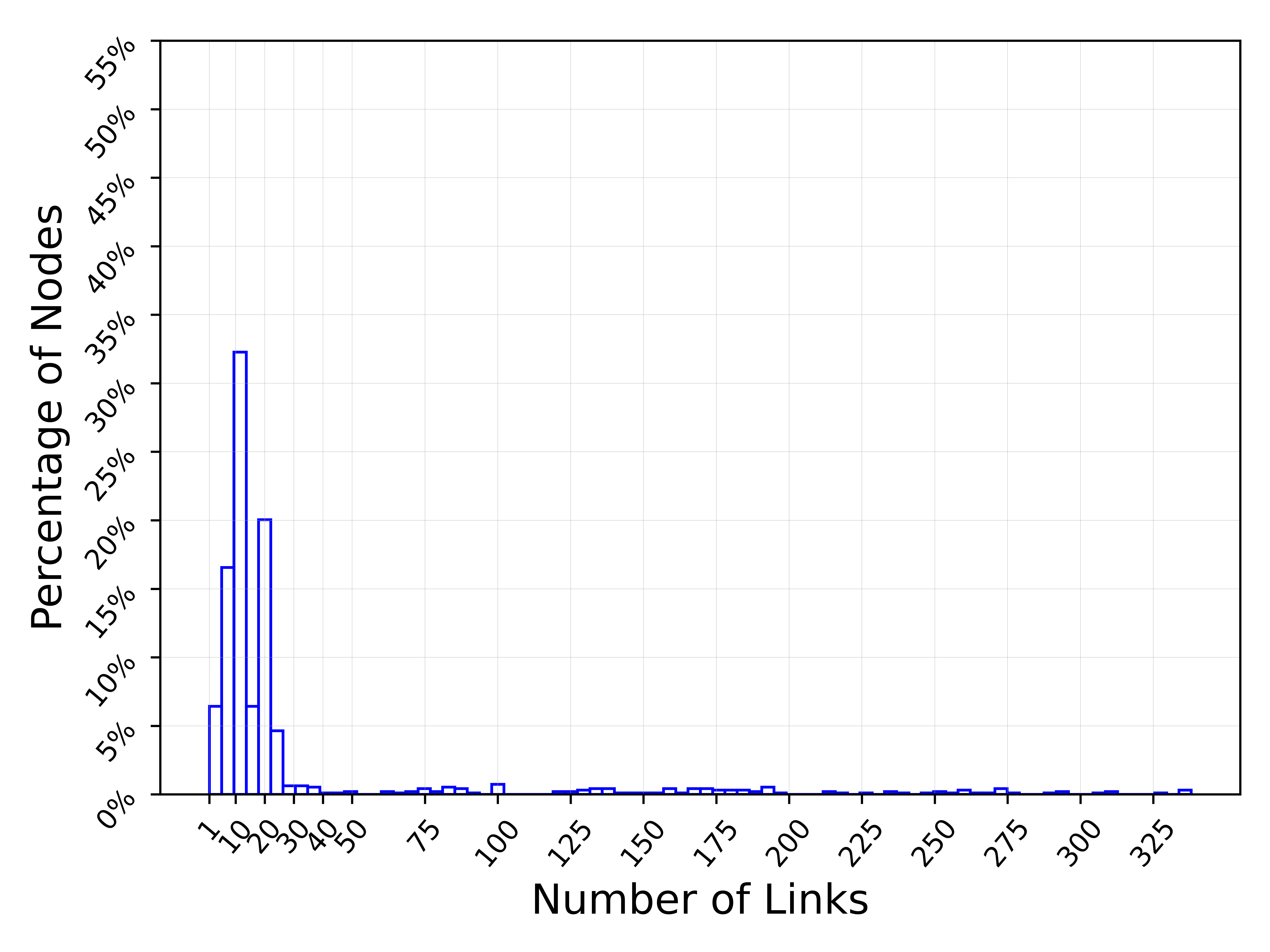}
        \caption[]%
        {{\small Total connection distribution. }}    
        \label{fig:deg-hist}
    \end{subfigure}\hfill
    \begin{subfigure}[t]{0.95\columnwidth}   
        \centering 
        \includegraphics[width=\textwidth]{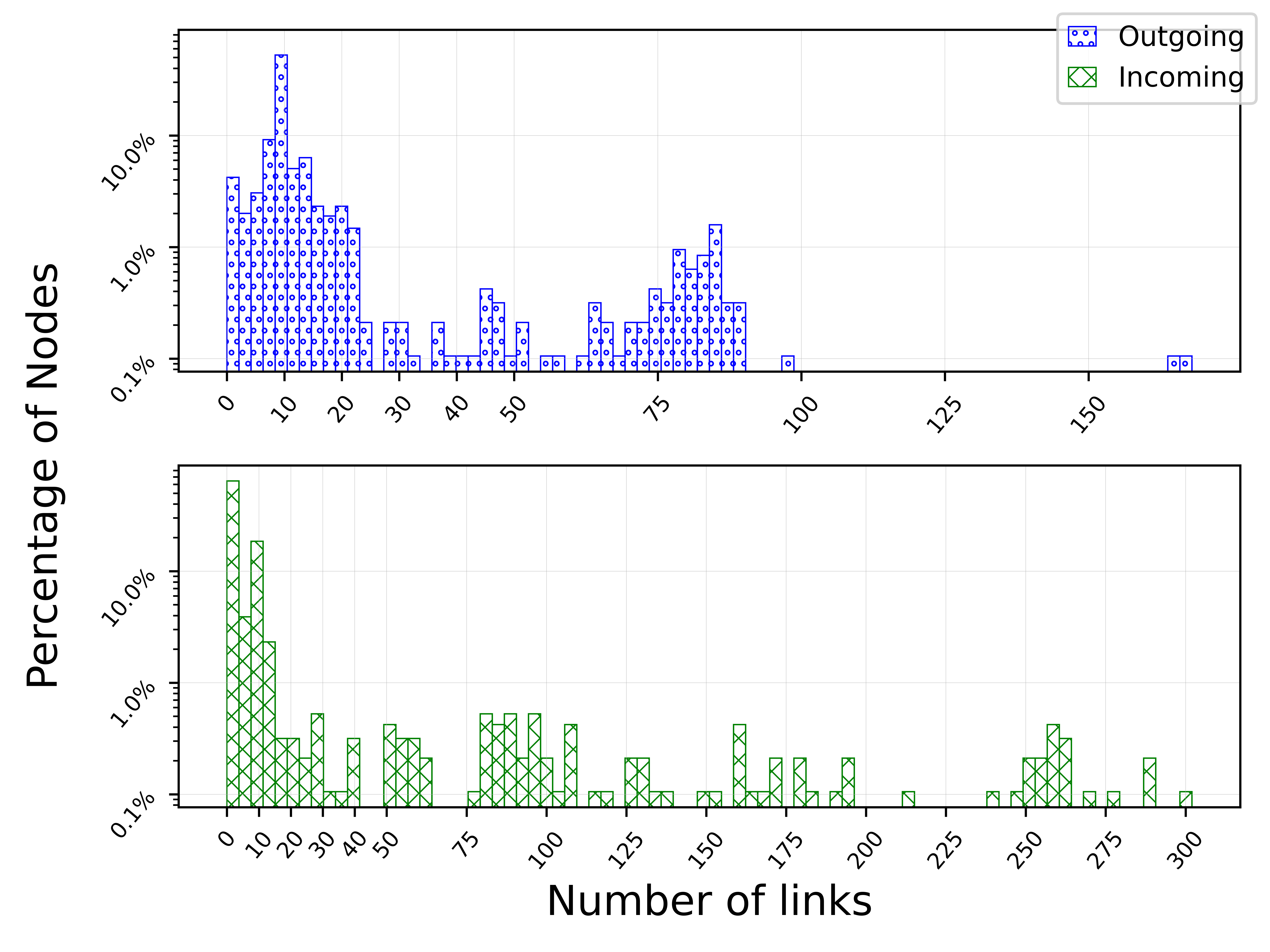}
        \caption[]%
        {{\small Outgoing (top) and Incoming (bottom) degree distribution. }  }  
        \label{fig:deg-hist-sep}
    \end{subfigure}
    \caption[  ]
    {Illustrative node degree distribution.} 
    \label{fig:deg_dist}
    \centering
\end{figure*}

\begin{table}[h]
\centering
\caption{\label{table:xrp-summary} Basic XRP network properties.}
\begin{tabular}{l|c|c}
                                  & \textbf{Mean}  & \textbf{STD}  \\\hline
\textbf{Nodes}                              & 948.53          &  18.54       \\\hline
\textbf{Edges}                              & 15010.26        &  508.92      \\\hline
\textbf{In-Degree}                          & 15.82           &  45.62       \\\hline
\textbf{Out-Degree}                         & 15.82           &  19.94       \\\hline
\textbf{Connected Component}                & 1               &  0.00        \\\hline
\textbf{Assortativity}                      & -0.48           &  0.02       \\\hline
\textbf{Global Clustering Coefficient}      & 0.76            &  0.02       \\\hline
\textbf{Density}                            & 0.03            &  0.00       \\\hline
\textbf{Avg. Shortest Path}                 & 2.31            &  0.03       \\\hline
\textbf{Diameter}                           & 5.1             &  0.33         \\\hline
\end{tabular}
\end{table}

We summarize the basic properties of the XRP Ledger Network in Table~\ref{table:xrp-summary}. 

\paragraph{Size} The network is relatively small. We observed 948 nodes and 15,010 links on average. In comparison, Bitcoin has 50,000 nodes and Ethereum 12,000 nodes~\cite{paphitis2021}. We measured a fluctuation of 2\% in the total node count and 1\% in the edge count.

\paragraph{In\&Out Degrees} Each outgoing connection corresponds to an incoming one, and the nodes report only the active links (not the potential ones). Therefore, the means of incoming and outgoing degrees are equal. The standard deviation of incoming connections is 45.62. This is more than two times greater than that of outgoing ones. The difference in deviations suggests that some nodes in the network accept significantly more incoming connections than others.

\paragraph{Connected Components} A network is consistently connected when, at any point in time, there is a path between two nodes in the network. XRP Ledger is consistently connected, as indicated by the single connected component and zero-value standard deviation. However, this may also be due to the nature of the crawler. The crawler can only discover the nodes that are members of the same connected component as the initial entry node. However, any node not connected to the core could not participate in the Ledger. 

\paragraph{Network Density} Network Density is the proportion of the \textit{possible} and \textit{actual} connections in the network. Higher values indicate a denser network. In dense networks, messages have a lower propagation delay but at the cost of increased redundancy~\cite{barabasi2016network}. The XRP Ledger network has a density of 0.03. In comparison, the density of Bitcoin and Ethereum are 0.002 and 0.0006, respectively~\cite{paphitis2021}. 

\paragraph{Clustering Coefficient} The Clustering Coefficient quantifies the node's tendency to form tightly knit groups with high-density ties. The Global Clustering coefficient is 0.78. The low average shortest path and high clustering coefficient of XRP Ledger Network suggest that it may exhibit the small-world property. 

\begin{figure*}[htbp]
    \centering
    \begin{subfigure}[t]{0.95\columnwidth}
        \centering 
        \includegraphics[width=\textwidth]{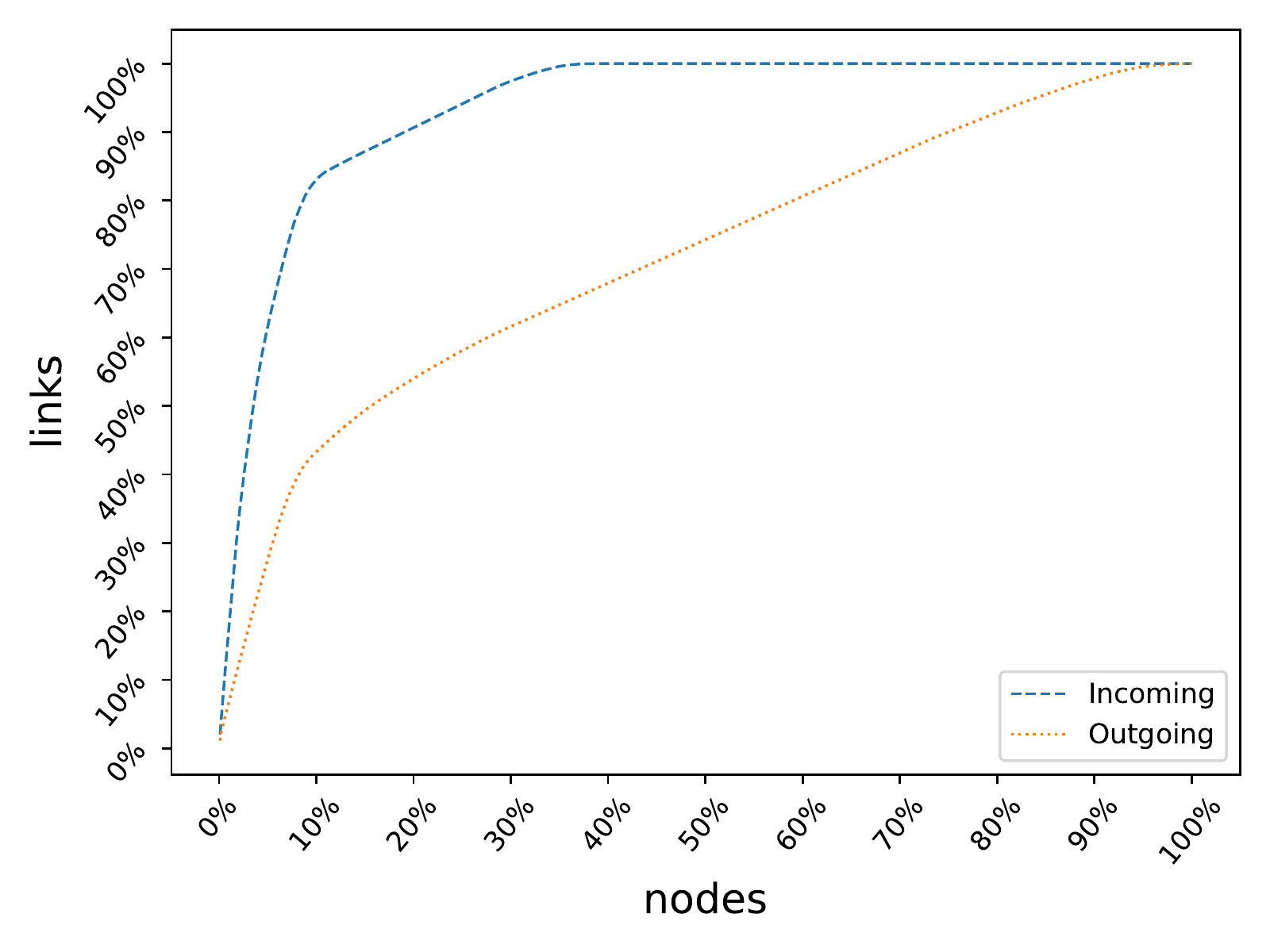}
        \caption[]%
        {{\small Connection Distribution amongst the nodes. }}    
        \label{fig:cumsum-perc}
    \end{subfigure}\hfill
    \begin{subfigure}[t]{0.95\columnwidth}   
        \centering 
        \includegraphics[width=\textwidth]{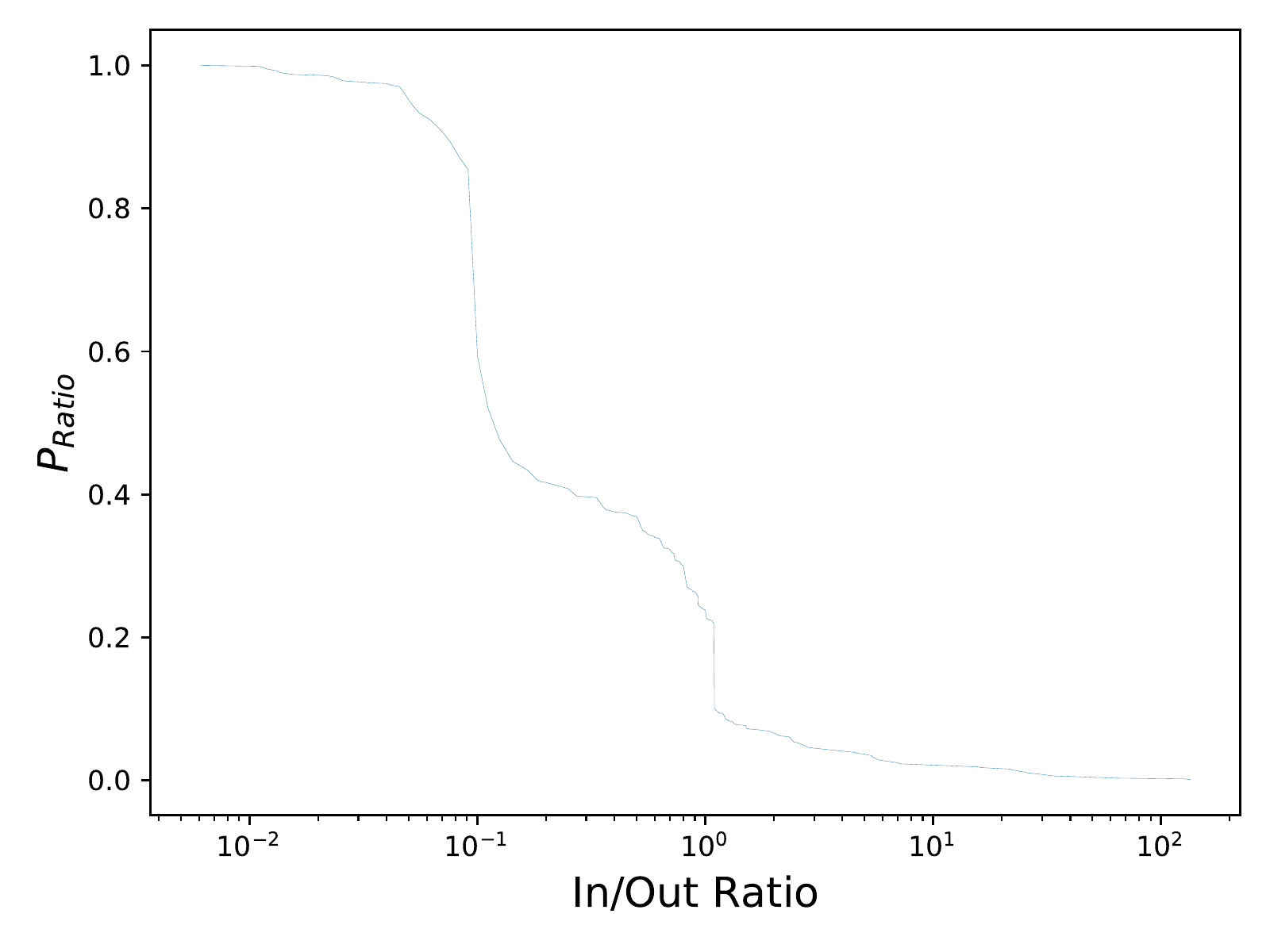}
        \caption[]%
        {{\small In/Out Degree Ratio.}}    
        \label{fig:ratio}
    \end{subfigure}
    \caption[  ]
    {Representative network properties.} 
    \label{fig:single-fig}
\end{figure*}

\subsection{Single Network}
\label{sec:single-network}

We conducted an in-depth analysis of a single XRP Ledger network snapshot. We selected the sample whose node and edge counts are the closest to the mean of the dataset. In the remainder of this Section, we discuss our findings.

\paragraph{Degree Distribution}
\label{sec:deg-dist}

The network degree distribution impacts many of its properties, such as message propagation delay and the resilience of the network~\cite{barabasi2016network}. Random networks have binomial degree distributions, whereas real-world networks contain a small number of highly connected nodes that cannot be accounted for by random models~\cite{Broido}.

In Figure~\ref{fig:deg-hist}, we illustrate the percentage of nodes (y-axis) with a given number of combined incoming and outgoing connections (x-axis). The distribution's shape is similar to a gamma distribution with a long right tail. The majority of nodes,  32.5\%,  have between 10 and 15 connections. At the tail end, nodes have over 325 peers, six times more than nodes at the beginning of the tail.

In Figure~\ref{fig:deg-hist-sep}, we show separated incoming and outgoing distributions. The upper plot depicts the outgoing connections. The majority of nodes establish between 1 and 22 connections. The largest bin holds 50\% of all the nodes, with ten peers. The spike reflects the default \textit{rippled} configuration. At the time of the data collection, the default number of outgoing connections was 10. This value was since updated to 21~\cite{rippled_default_peer}. Other nodes connect to between 22 and 90 peers. We also found three outliers; Two nodes with well over 150 and one with just under 100 connections. 

In the lower plot, we depict the incoming connection distribution. The first bin contains 60\% of nodes without incoming connections. There is no incentive to accept connections, but there is a server maintenance cost. Therefore, the majority of servers only establish outgoing connections. In addition, the first bin may also include validators. By default, for security reasons, they do not accept incoming connections.

The second largest group represents 19\% of nodes with between 9 and 11 connections. The remaining bins contain 11\% of nodes. These account for the vast majority of the incoming connections in the network. Nodes with around 150 incoming connections are the hubs.

In Figure~\ref{fig:cumsum-perc}, we illustrate the cumulative sum of incoming and outgoing connections. There are two outgoing connection groups. The first group, indicated by the exponential portion of the curve, holds nodes whose out-degree is above the mean. It contains \~15\% of the nodes that account for \~50\% of all connections. The second group, indicated by the linear portion of the curve, holds the remaining 85\% of the nodes. Finally, a deeper inspection revealed two outlier nodes with over 150 outgoing connections. 

We similarly grouped the incoming connections. The first group, indicated by the sharp spike of the curve, dominates the overall network connectivity. It contains 11\% of nodes with an in-degree above the mean. These nodes account for 85\% of incoming connections. The second group, depicted by the short linear portion of the curve, contains 27\% of nodes. They account for approx. 15\% of the incoming links. The final group, reflected by the plateau, holds nodes without incoming connections and accounts for the remaining 62\% of nodes.

The incoming and outgoing connection distributions are heavy-tailed. However, they seem to have different shapes. We discuss which model best describes these distributions in Section~\ref{sec:scale-free}. We also observe that a small subset of nodes holds the majority of connections. Our findings suggest that the network has a group of authoritative nodes.

\begin{figure*}[htbp]
    \centering
    \begin{minipage}[t]{0.95\columnwidth}
        \centering
       \includegraphics[width=\textwidth]{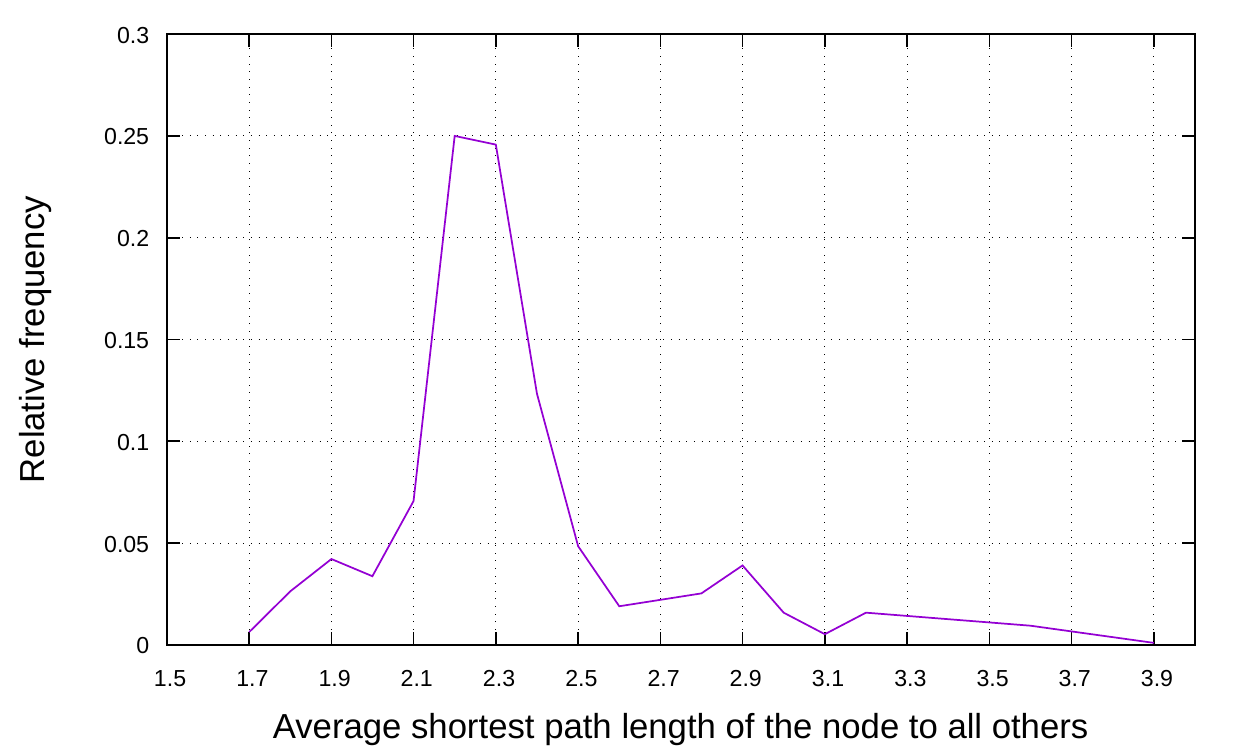}
        \caption{Distribution of the average shortest path length.} 
        \label{fig:dist-dist}
    \end{minipage}\hfill
    \begin{minipage}[t]{0.95\columnwidth}
        \centering
        \includegraphics[width=\textwidth]{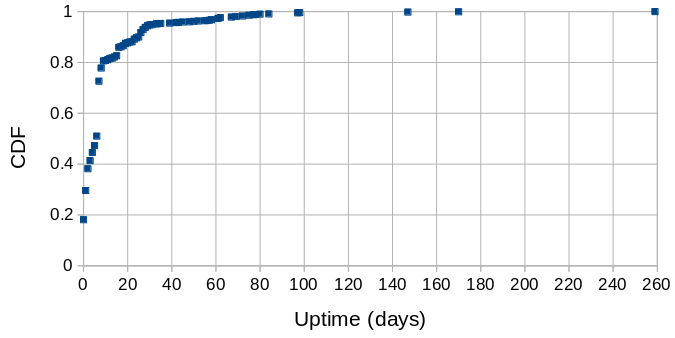}
        \caption{Node uptime CDF.}
        \label{fig:dist-uptime}
    \end{minipage}
\end{figure*}

\paragraph{Scale-Free Property}
\label{sec:scale-free}

Across scientific domains, it is often claimed that real-world networks are scale-free. Details vary, but in general, a network is scale-free, when nodes with degree $k$ follow a power-law distribution $k^{-\alpha}$, where $\alpha$ is the scaling criterion $\alpha > 1$. However, other versions of this hypothesis have stronger restrictions, e.g. $2 < \alpha < 3$~\cite{barabasi2016network}. Cohen \emph{et al. } show that scale-free networks are highly resilient to random attacks but are vulnerable to targeted attacks~\cite{Cohen2001}. Therefore, it is important to understand the type of degree distribution.

We used the \textit{fitter}~\cite{fitter} Python library to find the most accurate model. We used the Akaike Information Criterion (AIC) and Bayesian Information Criterion (BIC) to determine the quality of a fit. A lower AIC or BIC value indicates a better fit. The analysis in Section~\ref{sec:deg-dist} revealed that the in and out degrees are likely to have different distributions. Therefore we modelled the in, out, and combined distributions separately. We found that the in-degree distribution was best captured by the \textit{power-law} distribution, with $\alpha ~ 1.2$. On the other hand, the out-degree was best described by the \textit{generalized normal distribution}, with a heavy, long tail. Likewise, \textit{generalized normal distribution} fits the overall degree distribution the best.

The ubiquity of scale-free networks in the real world has been questioned~\cite{Broido}. Therefore, we avoid claiming that the XRP network is scale-free, as a deeper analysis is required. However, our findings indicate that the XRP network is not random. Furthermore, the power-law distribution fit of the in-degree offers further evidence that the network relies on a subset of nodes for its connectivity.

\paragraph{Small-World Property}
The well-studied small-world property indicates that a short path connects any two nodes in the network~\cite{barabasi2016network}. An average shortest path $\textit{l}$ is short when $l \approx \frac{ln N}{ln \langle k \rangle }$, where N is the size of the network, and $\langle k \rangle$ is the average degree.

Manfred Kochen and Ithiel de Sola Pool~\cite{de1978contacts} formalized the effect. Which was later popularized by the well-known Milgram experiment that inspired the \textit{six degrees of separation} phrase. 

Network G is said to be small-world if it has a similar average shortest path length but a greater clustering coefficient than an equivalent random graph. Two graphs are equivalent when they have an equal number of nodes. Let $L_g$ be the average shortest path length of $G$ and $C_g$ its clustering coefficient. Equivalent properties for a random graph are $L_{rand}$ and $C_{rand}$. Network G is said to be small-world if $L_g \geq L_{rand}$ and $C_g >> C_{rand}$. 

A quantitative measure of small-worldness is expressed as follows: $\gamma_g = \frac{C_g}{C_{rand}}$ and $\lambda_g = \frac{L_g}{L_{rand}}$, where $\gamma_g$ is the clustering coefficient ratio and $\lambda_g$ is the average shortest path ratio of network G and an equivalent random graph. Then measure of small-worldness is expressed as $S = \frac{\gamma_g}{\lambda_g}$. A network is considered small-world when $S > 1$~\cite{Humphries2008}.

We used the Erdős–Rényi (ER) model to generate random graphs. To ensure the robustness of the small-worldness calculation, we used Monte Carlo sampling of 1000 equivalent ER graphs. We measured $S = 8.3$ for the XRP Network. We, therefore, conclude that the XRP network has the small-world property.

\paragraph{In/Out Degree Analysis}

Link analysis is a method to identify authoritative nodes in a network~\cite{Kleinberg, Mislove}. We use it to identify selfish nodes that do not reciprocate the connections they establish by accepting incoming links.  

We express the link ratio as $\lambda = \frac{In+1}{Out+1}$, a ratio between incoming and outgoing number of connections. All degrees are incremented by $1$ to account for no incoming or outgoing connections. A high ratio $\lambda > 1$ suggests that a node is altruistic - it establishes more incoming connections than outgoing ones. Conversely, $\lambda < 1$ indicates nodes that consume more connectivity than they provide.

We illustrate the ratio distribution in Figure~\ref{fig:ratio}. We observe that 15\% of nodes have $\lambda << 1$. Interestingly, we find that a significant percentage of nodes have a $\lambda = 0.09$. These nodes use the default \textit{rippled} configuration, with ten outgoing and zero incoming connections. In contrast, only about 10\% of nodes have more incoming than outgoing connections, and only 3\% $\lambda >> 1$.

There are no direct incentives to participate in the XRP network. However, running a node that accepts incoming connections requires significant investment. Such a server has to be reliable and available. Therefore, we see that most nodes connect to the network as \textit{consumers}, and only relatively few behave altruistically. In the next section, we discuss the preference of nodes to connect to other similar nodes.

\paragraph{Degree Correlation}

The \textit{degree correlation} captures the node's preference to form connections with others that are similar in some way~\cite{barabasi2016network}. In the context of this study, we consider similarity in terms of node degree. A network is \textit{assortative} when nodes tend to connect to others with a similar degree. In a \textit{disassortative} network, small-degree nodes prefer to link with high-degree nodes, and hubs tend to avoid each other. Finally, a network is considered \textit{neutral} when the wiring between the nodes is random. 

The \textit{degree correlation} impacts the robustness of a network~\cite{Vazquez2003}. In an assortative network, node removal causes little fragmentation, as high-degree nodes form a core group and are redundant. In contrast, disassortative networks are easier to fragment~\cite{barabasi2016network}. High-degree nodes connect to many small-degree nodes, forming a hub-and-spoke structure. The small-degree nodes become disconnected once a high-degree node fails.

Degree \textit{correlation coefficient} \textit{r} characterizes degree correlation using a single number \textit{r}~\cite{Newman2003}. In general it varies between $-1 \leq r \geq 1$~\cite{barabasi2016network}. For $r > 0$ the network is assortative, for $r = 0$ the network is neutral, and for $r < 0$ the network is disassortative. The degree correlation coefficient for XRP Ledger is $-0.48$. In comparison, the degree correlation of an equivalent ER network is zero. We conclude that the XRP Ledger network is disassortative. It has a hub-and-spoke network structure and may be vulnerable to targeted attacks.

\begin{figure*}[htbp]
    \centering
    \begin{subfigure}[t]{0.95\columnwidth}  
        \centering 
        \includegraphics[width=\textwidth]{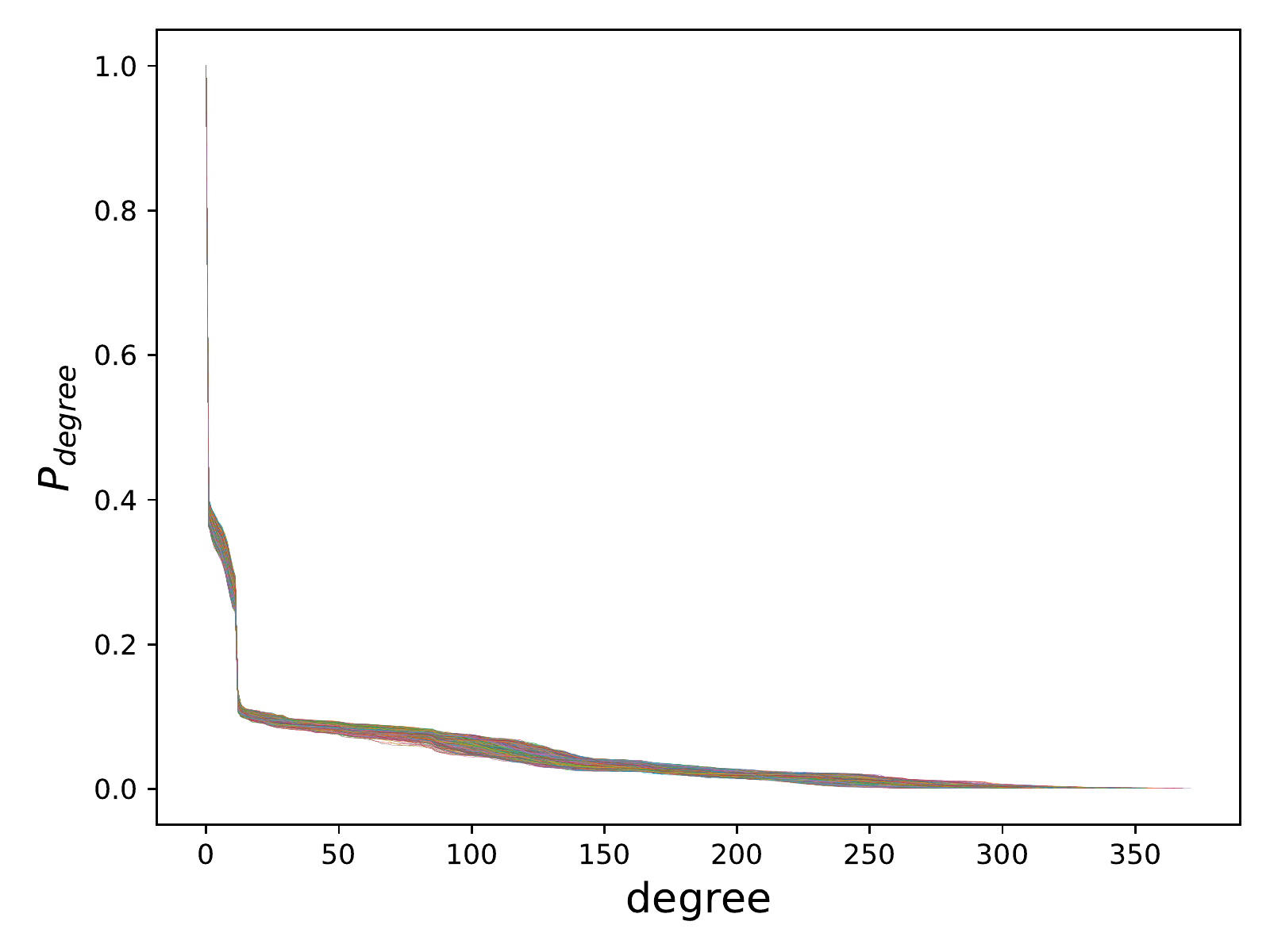}
        \caption[]%
        {{\small Incoming connections.}}    
        \label{fig:comb-in-deg-ccdf}
    \end{subfigure}\hfill
    \begin{subfigure}[t]{0.95\columnwidth}   
        \centering 
        \includegraphics[width=\textwidth]{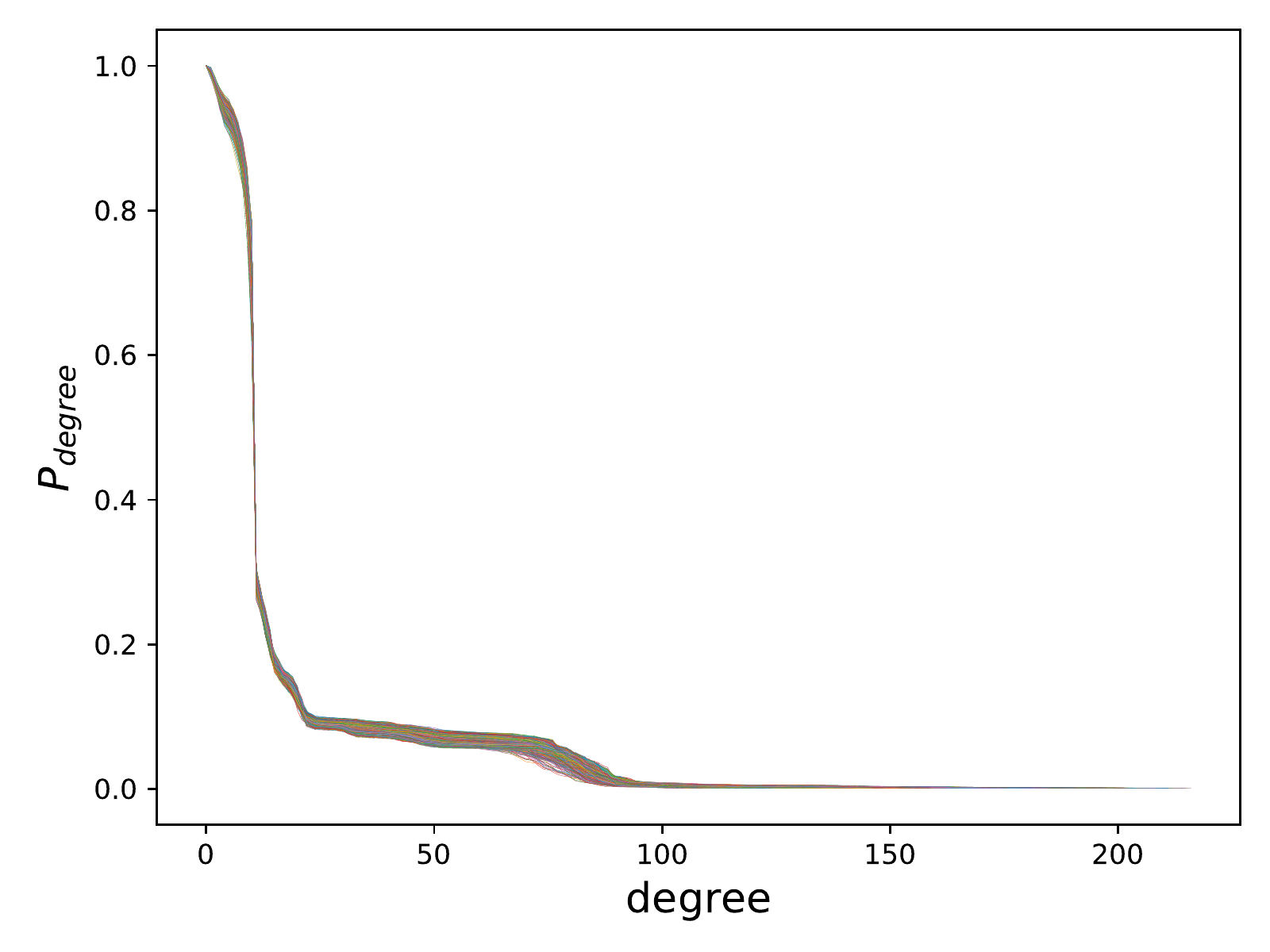}
        \caption[]%
        {{\small Outgoing connections.}}    
        \label{fig:comb-out-deg-ccdf}
    \end{subfigure}
    \caption[  ]
    {Temporal complimentary cumulative degree distribution.} 
    \label{fig:temp-degree-ccdf}
\end{figure*}

\paragraph{Average Shortest Path Distribution}
\label{sec:dist-dist}

The average shortest path is the mean of all shortest paths from a node to every other node in the network. We show the distribution of these distances in Figure~\ref{fig:dist-dist}. 

The X-Axis is the average path length (in hops), rounded to the tenth. The Y-Axis is the percentage of nodes with the given path length.   We observe that around 50\% of the nodes have an average distance between 2.2 and 2.3 hops. The distribution has a bell-like shape with a long tail towards longer distances. IP networks have a similar distribution, although the average path length is around nine hops~\cite{1033345}. 

XRP Ledger uses broadcasting to propagate messages in the network. The shortest path distribution suggests that the network's topology assists in timely message delivery.

\paragraph{Node Distribution over Autonomous Systems}
\label{sec:dist-as}

The Autonomous System (AS) number is a 32-bit unique identifier. It represents a collection of IP networks administered by a single entity. We used the AS number to compute the node distribution per AS. In the remainder of this section, we discuss our findings.

Most systems contain only a handful of nodes, while a few AS have a large number of nodes. Around 18\% of discovered nodes did not reveal their IP addresses. Therefore, we do not know their AS details.

We split AS details across two tables to illustrate the heavy-tailed nature of node distribution between the AS. Table~\ref{table:top-as} shows the top ten AS by the number of nodes. These systems, owned by the largest cloud service providers (Amazon, Google, Microsoft), hold nearly 62\% of the XRP Ledger nodes. Routing failures in one or more of the dominant Autonomous Systems may disrupt XRP Ledger operations. Therefore, a concentration of nodes may represent a weakness for the ledger.
 
The remaining 38\% of the nodes are evenly distributed over 117 AS. Table~\ref{table:node-as} shows the number of AS possessing a given number of nodes. We observe that 84 AS only host one XRP node. As the node count per AS increases, the number of autonomous systems rapidly approaches 1.

\begin{table}[h]
\centering
\caption{\label{table:top-as}AS with the highest number of nodes.}
\begin{tabular}{c|l|l|c}
\hline
\textbf{Rank} & \textbf{AS number} & \textbf{AS name}	& \textbf{XRP nodes} \\
\hline
1 & 16509	& Amazon.com & 177 \\
2 & 24940	& Hetzner Online & 115 \\
3 & 14618	& Amazon.com & 71 \\
4 & 8987	& Amazon DS Ireland & 70 \\
5 & 396982	& Google & 52 \\
6 & 8075	& Microsoft Corporation & 25 \\
7 & 16276	& OVH & 23 \\
8 & 14061	& DigitalOcean & 18 \\
9 & 38895	& Amazon.com & 18 \\
10 & 134963	& Alibaba.com Singapore & 17 \\
\hline
\end{tabular}
\end{table}

\begin{table}[h]
\centering

\caption{\label{table:node-as} Number of AS with a given number of nodes.}
\begin{tabular}{c|c}
\hline
\textbf{Nodes per AS} & \textbf{Total number of AS} \\
\hline
1 & 84\\
2 & 13\\
3 & 8\\
4 & 6\\
5 & 3\\
6 & 1\\
8 & 2\\
\hline
\end{tabular}
\end{table}

\paragraph{Node Uptime Distribution}
\label{sec:uptime}
The \textit{rippled} software reports the number of seconds (uptime) it has been running. We plot the cumulative distribution function (CDF) of the reported uptime in Figure~\ref{fig:dist-uptime}. The X-Axis depicts the uptime in days, rounded to the closest hour.

The average uptime is 9.7 days, with a standard deviation of 18.4 days. Just under 18\% of nodes reported uptime of fewer than 12 hours, whereas the oldest node was running for 259 days. Approximately 18\% of nodes reported an uptime between 20 and 60 days, and 2\% were running for up to 80 days. We observed only a handful of nodes older than 100 days.

\begin{figure*}[htbp]
    \centering
    \begin{minipage}[t]{0.95\columnwidth}
        \centering
        \includegraphics[width=\textwidth]{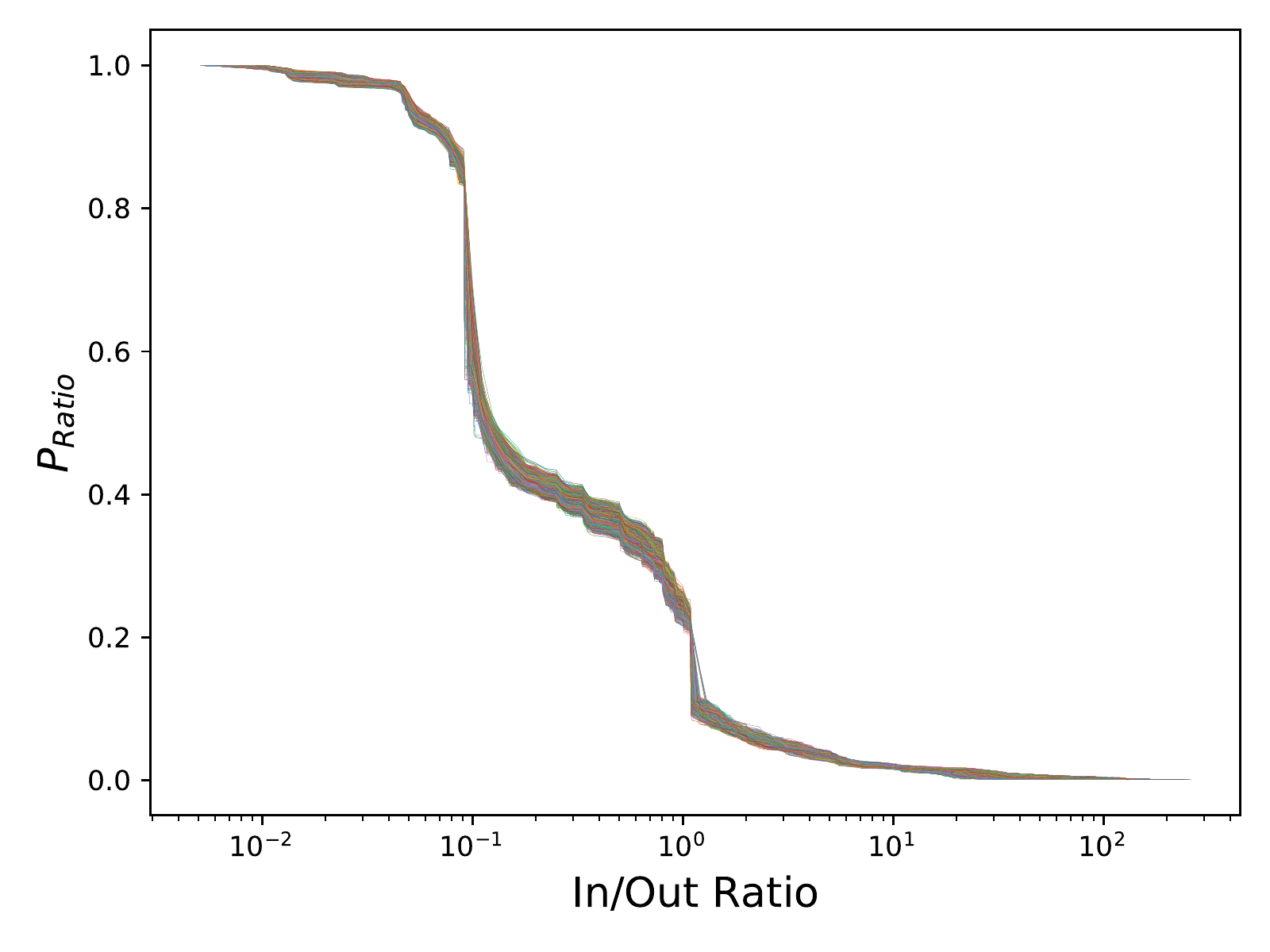}
        \caption{Temporal in/out degree ratio.} 
    \label{fig:temp-deg-ratio}
    \end{minipage}\hfill
    \begin{minipage}[t]{0.95\columnwidth}
        \includegraphics[width=\textwidth]{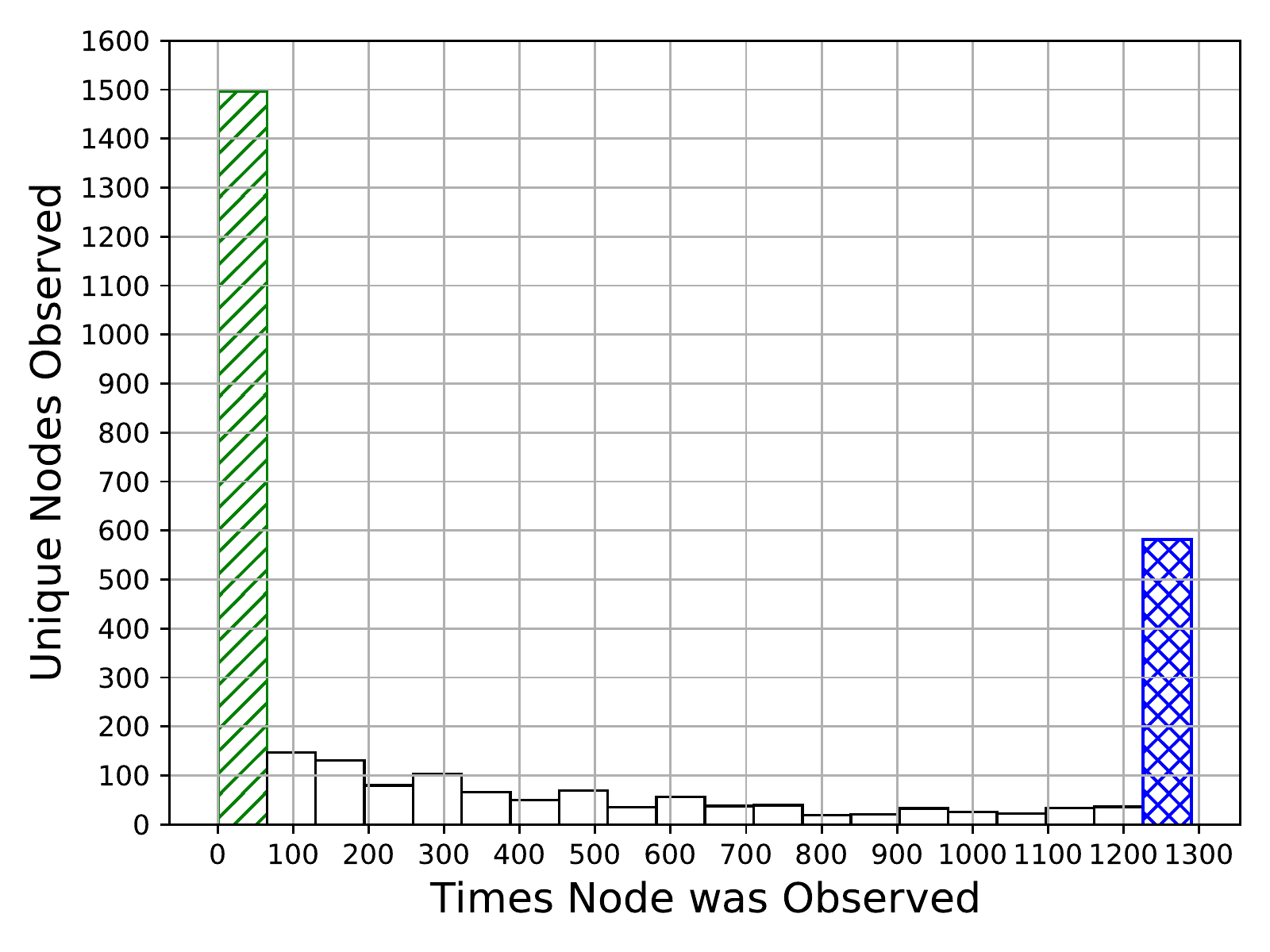}
        \caption{Number of times a node was observed in a network.} 
        \label{fig:temp-churn}
    \end{minipage}
\end{figure*}

\section{Temporal Analysis}
\label{sec:temporal-analysis}

In this section, we discuss the evolution of the network over time. We begin with a summary of the temporally stable properties. We observe that the preference for small-degree nodes to connect to high-degree nodes remains constant over time. Likewise, the global clustering coefficient and average shortest path are stable. Furthermore, all network snapshots have the small-world property. These findings suggest that there were no significant disruptions in the network during the observation period.

The relatively small change in the network's size had a non-negligible effect on the average incoming and outgoing degree, as indicated by the standard deviation. We dedicate the rest of this section to discussing these changes.

\paragraph{Degree Distribution}

We illustrate the complementary cumulative distribution function (CCDF) of the node degrees in Figure~\ref{fig:temp-degree-ccdf}. Overall, both distributions have long tails, and their shapes remain stable. However, we see some variance over time in both figures, as indicated by the changing thickness of the plots.

We plot the CCDF of incoming connections in Figure~\ref{fig:comb-in-deg-ccdf}. Our first observation is that consistently \~60\% of nodes do not accept incoming connections. Likewise, we see little variance at the tail-end of the spectrum. We see slightly more variance in nodes close to the mean and nodes with a degree between 250 and 300. We observe the largest variance in nodes with in-degree between 50-150. Our observations suggest that the nodes at the ends of the distribution are saturated. They cannot accept new peers. Therefore, nodes in the middle of the distribution handle the new connections to the network. Furthermore, the majority of new nodes do not accept incoming links.

We depicted the CCDF of the outgoing connections in Figure\ref{fig:comb-out-deg-ccdf}. The long, thin tail of the distribution suggests the existence of a few stable nodes with a high number of outgoing connections. We see a much higher variance in the group of nodes with an out-degree between 50 and 100. Finally, the majority of new nodes had an out-degree under the mean.

Two versions of the \textit{rippled} software came out during the data collection. Some observed variances may be explained by nodes leaving the network to update their version. However, overall the network has a stable member group.

\paragraph{In/Out Degree Analysis}

In Figure~\ref{fig:temp-deg-ratio}, we display the degree ratio plot for all captured snapshots. We observe little change in the overall degree ratio. The majority of nodes establish more outgoing than incoming connections. Only \~10\% of nodes establish more incoming than outgoing connections.

The lack of change in the shape of the curve confirms our initial observation that nodes do not reciprocate the connections they consume.

\paragraph{Membership Stability}

Over the collection period, we discovered 3,000 unique nodes. In Figure~\ref{fig:temp-churn}, we outline the lifespan of these nodes. The green, striped bar indicates nodes with the shortest lifespan. These nodes were present in around 5\% of all network snapshots. On the other side of the plot, the blue crossed bar represents the most stable nodes. They were present in at least 95\% of all the snapshots. The remaining 1/5th of the nodes have a gradually decreasing lifespan.

The fully present nodes have an average in-degree of 26.1. In comparison, the nodes we observed in the 5\% of snapshots have an average in-degree of 16. The difference between the values suggests that the fully present nodes are the ones that form the network backbone, which we discovered in Section~\ref{sec:deg-dist}.

We further analyzed the presence of the top 10\% of the highest in-degree nodes in the network over time. The group of the first network snapshot contains 95 nodes. The last network snapshot group holds 98 nodes. However, 23 nodes or 24\% from the first group are not present in the second group. Four nodes changed their IDs but had the old IP addresses and similar degree profiles. However, we did not find the other 19.

\paragraph{Node Uptime Over Time}

\begin{figure}
        \centering
        \includegraphics[width=0.95\columnwidth]{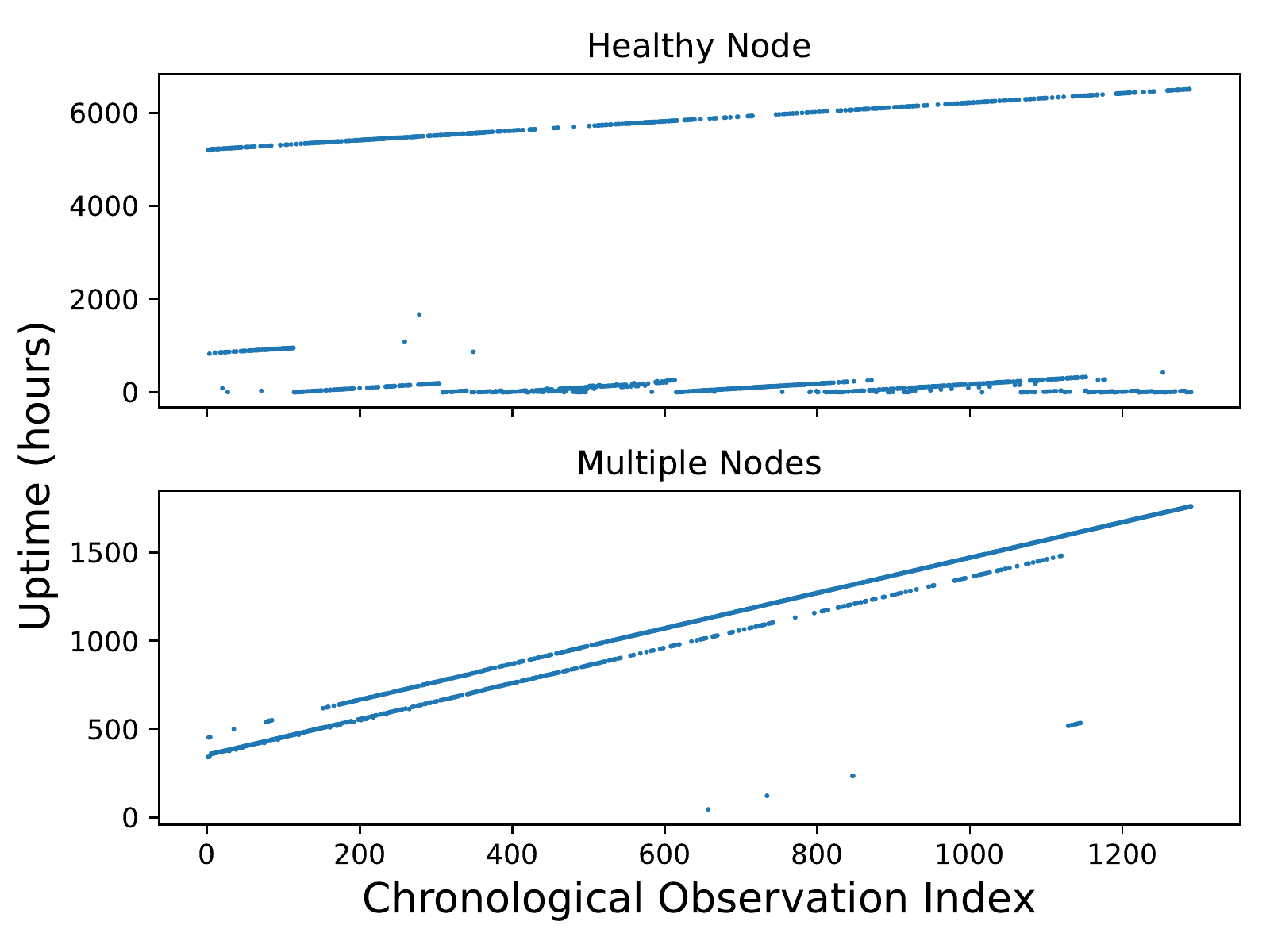}
        \caption{Illustrative uptime of representative nodes.} 
        \label{fig:temp-uptime}
\end{figure}\hfill

We measured the uptime of the 410 nodes present during every network crawl. Figure~\ref{fig:temp-uptime} presents two illustrative examples of the observed uptime. We limit our selection to the most representative nodes, in which we can observe clear patterns.
 
The top graph depicts the uptime graph of the 259 days old node we discussed in Section~\ref{sec:uptime}. There are two distinct features in the figure. The top line indicates that a server was running smoothly for the obersvation period. In contrast, the bottom feature suggests a server started and failed multiple times. These features suggest two separate instances of \textit{rippled} running behind a single IP address. 

The bottom plot offers a better illustration of two servers behind one IP address. There are two parallel lines of similar length. When we query the server uptime, we receive a response from one of the two servers. The fragmentation in the lines is information gaps caused by a different server handling the uptime request.

In all plots, we notice inexplicable uptime values. These could suggest that a new \textit{rippled} instance started or, more worryingly, uptime reporting issues. However, we leave the study of these observations for future work.

\section{Conclusion \& Future Work}
\label{sec:discuss}

A decentralized peer-to-peer overlay network forms the backbone of the XRP Ledger. In this paper, we provided an in-depth analysis of the graph-theoretic properties of this overlay.

We use a publicly available crawler to capture 1,290 snapshots of the underlying overlay network over two months. We find that it is significantly smaller than other blockchain overlay networks. The nodes are connected via short paths and are tightly clustered. Furthermore, the clusters tend to have a hub-and-spoke structure, as shown by the high assortativity of the network. Unlike other blockchain overlay networks, XRP has a small-world topology.

XRP does share some similarities with other blockchains. The network degree distribution has an exponential-like shape. We did not find conclusive evidence that it is scale-free. However, like other blockchains, the topology is not random.

Overall, the size of the network is consistent over time. However, we captured a significant amount of churn. Given these observations, we suspect that many nodes join the network to conduct their business and leave shortly.

The XRP overlay network may be vulnerable to targeted attacks. We discovered the existence of a small subset of influential nodes that provide the backbone of the network connectivity. Furthermore, a malicious actor can use the publicly available topology to identify these nodes.

We revealed a vast disparity between nodes that accept incoming connections and nodes that do not. Furthermore, link analysis showed that many nodes do not accept incoming connections. These nodes increase the dependence on the influential nodes. Thus, contributing to the network centralization. We suspect that a lack of financial incentive contributes to this behavior, as there are significant costs associated with running a reliable node. Natural centralization is a common problem in decentralized peer-to-peer networks\cite{karakaya2009free}\cite{hughes2005free}. A common solution is to introduce communal incentives or mandatory behavior. 

Our results raise further questions about the security and vulnerability of the XRP Ledger. Research works~\cite{Cohen2001}~\cite{Peng2016} show that networks with a long-tail degree distribution are susceptible to targeted attacks. For future work, we intend to evaluate the resilience of the XRP network to random and targeted attacks, and to identify mitigation strategies.

\section*{Acknowledgment}
This work was supported by Ripple UBRI\footnote{https://ubri.ripple.com/}.

\bibliographystyle{./ACM-Reference-Format}
\bibliography{./bibliography} 

\end{document}